\documentclass[12pt]{article}
\usepackage[dvips]{graphicx}
\newcommand{\be}{\begin{eqnarray} }
\newcommand{\ee}{\end{eqnarray} }
\newcommand{\beq}{\begin{equation} }
\newcommand{\eeq}{\end{equation} }

\begin{document}
\begin{center}
{\Large
Comment on polarized quark distributions extracted from SIDIS
experiments  
}
\vskip 1.5cm
{ \large A.N.~Sissakian}
{ \large O.Yu.~Shevchenko},
{ \large O.N.~Ivanov}
\\
\vspace{1cm}
{\it Joint Institute for Nuclear Research\\
Dubna, Moscow region 141980, Russia}\\

\end{center}

\begin{abstract}

The results of  SIDIS experiments concerning the first moments of the
polarized quark distributions are considered. 
The possible reasons of the deviation from the fundamental restrictions
such as the Bjorken sum rule and the  ways to properly 
improve
the analysis of measured SIDIS asymmetries are discussed. 
The possibility of broken polarized sea scenario is analysed. 
\begin{flushleft}
PACS number(s): 13.60.Hb, 12.38-t, 13.30-a, 13.88.+e
\end{flushleft}

\end{abstract}
\vskip 0.5cm

The extraction of the polarized quark and gluon densities is the main task of 
the SIDIS experiments with the polarized beam and target.
Of a special importance for the modern SIDIS experiments are the questions of strange
quark 
and gluon contributions 
to the nucleon spin, and, also the sea quark share as well as the possibility of broken sea scenario. 
Indeed,
it is known [1] that the unpolarized sea of light quarks is essentially asymmetric, and, thus, 
the question arises: does the analogous situation occurs in the polarized case,
i.e.  whether the polarized density $\Delta\bar u$ is equal to $\Delta\bar d$  or not. 

The crucial tests for the polarized quark distributions extracted from the SIDIS data 
are the sum rules dictated by $SU_f(2)$ and $SU_f(3)$ 
symmetries. While $SU_f(3)$ symmetry (and, as a consequence, the respective sum rule)
is rather approximate (see, for example [2] and refs. therein), $SU_f(2)$ symmetry
may  be regarded  as almost exact as well as the respective sum rule--Bjorken sum rule (BSR).

Let us remind that the Bjorken sum rule written in terms of the first moments of the structure functions 
$\Gamma_1^p(Q^2)\equiv\int_0^1dxg_1^p(x,Q^2)$ 
and $\Gamma_1^n(Q^2)\equiv\int_0^1dxg_1^n(x,Q^2)$  contains $Q^2$ dependent quantity $C_1^{NS}$ 
in the right-hand 
side\footnote{See, for example, excellent theoretical overview in [3] and references therein. 
The $O(\alpha_s^3)$ correction for $C_1^{NS}$ was calculated in [4], and,
$O(\alpha_s^4)$ correction was estimated in [5].
}:
\be
\Gamma_1^p&-&\Gamma_1^n=\frac{1}{6}\left|\frac{g_A}{g_V}\right|C_1^{NS}(Q^2), \\
C_1^{NS}&=&1-\left(\frac{\alpha_s(Q^2)}{\pi}\right)-3.5833 
\left(\frac{\alpha_s(Q^2)}{\pi}\right)^2 \nonumber \\
&-&20.2153\left(\frac{\alpha_s(Q^2)}{\pi}\right)^3
-130\left(\frac{\alpha_s(Q^2)}{\pi}\right)^4+O(\alpha_s^5).
\ee
However (and this is of great importance for what follows),
the first moments of polarized quark distributions satisfy the respective form of the BSR 
without
$C_1^{NS}$ in the right-hand side irrespectively in which QCD order they are
extracted. Namely, the equivalent of BSR written in terms of polarized
quark distributions reads   
\be
\label{a3num}
&&\Delta q_3\equiv a_3 = (\Delta_1 u(Q^2)+\Delta_1\bar u(Q^2))-(\Delta_1 d(Q^2)+\Delta_1\bar d(Q^2))\nonumber\\
&&=\left |\frac{g_A}{g_V}\right |=F+D=1.2670\pm0.0035\quad \mbox{\it      in all QCD orders},
\ee
where 
the notation $\Delta_1 q\equiv\int_0^1dx \Delta q$ is used 
to distinguish the local in Bjorken $x$ polarized quark densities $\Delta q(x)$ 
and their first moments.

Notice that well known fact of nonrenormalizability (i.e., $Q^2$ independence) of 
the quantity $\Delta q_3$
directly follows from its definition  

\be
\label{a3}
\frac{s_\mu}{2}\Delta q_3=\langle ps|A_\mu^3|ps\rangle
\ee
 due to conservation\footnote{
It is important to remind that while the first moments
of the {\it nonsinglet}  densities $\Delta q_3$ ($\rm SU_f(2)$ symmetry) 
and $\Delta q_8$ ($\rm SU_f(3)$ symmetry) {\it must be }
conserved,
i.e. are independent of $Q^2$ (corresponding to the conservation
of the non-singlet axial-vector Cabibbo currents), 
the {\it singlet} axial charge,
$a_0(Q^2)$ depends on $Q^2$ because of the axial anomaly .}
of the flavour nonsinglet 
axial vector current $A_\mu^3$. 
This fact is also confirmed by the explicit NLO calculations of the  
respective nonsinglet anomaly dimension which is just zero [6], 
so that one has\footnote{Here the notation of Ref. [6] 
for the anomaly dimension is used.}

$$\frac{{\rm d} \Delta_1 q_3 }{ {\rm d}\ln(Q^2/\Lambda^2)}=
\frac{\alpha_s}{2\pi}\,\delta \gamma_{NS,\,\eta}^{(n)} {\Bigl |}_{n=1,\,\eta=-1}
 \,\Delta_1 q_3=0.
 $$

Let us analyse to what extent the results of the modern polarized SIDIS
experiments are in agreement with the sum rule predictions. 
Such detailed analysis with respect to the sum rule based on $ SU_f(3)$ symmetry 
$$
\Delta q_8 \equiv a_8 =3F-D,
$$
was performed in [2], so that we will concentrate here on the equivalent of BSR (3)
which, using that $\Delta q=\Delta q_V+\Delta\bar q$, may be
rewritten in the form convenient for analysis:
\be
\Delta_1\bar{u}-\Delta_1\bar{d}=\frac{1}{2}\left|
 \frac{g_A}{g_V}\right|-\frac{1}{2}
(\Delta_1 u_V-\Delta_1 d_V)\quad\mbox{\it   in all QCD orders}.
\ee

Let us first consider the SMC results [7].
SMC has performed two types of analysis on $\Delta q$, with broken and unbroken sea scenarios,
respectively. Unfortunately, the SMC analysis without the unbroken sea assumption     
suffers from too big errors\footnote{ Indeed, for $\Delta_1 \bar d$ the
Table 5 of Ref. [7] gives $\pm0.14$ and $\pm0.12$ for the statistical
and systematical errors, respectively.}
because the full number of measured asymmetries and achieved statistics 
were not quite sufficient to release the restriction
$\Delta \bar u =\Delta \bar d$.  
So, let us look at the SMC results for the first moments of polarized
quark distributions obtained within the unbroken sea scenario, where the respective
table of first moments looks as (see Table 5  of ref. [7] ) 
\vskip 0.5cm

\begin{tabular}{|c|c|c|c|c|}
\hline
 & & & & \\
$\Delta\bar u(x)=\Delta\bar d(x)$&x&0-0.003&0.003-0.7&0-1\\
\hline
 &$\Delta_1 u_V$&$0.04\pm0.04$&$0.73\pm0.10\pm0.07$&$0.77\pm0.10\pm0.08$\\
\hline 
 & $\Delta_1 d_V$&$-0.05\pm0.05$&$-0.47\pm0.14\pm0.08$&$-0.52\pm0.14\pm0.09$\\
\hline 
 & $x$ & $0.-0.003$ & $0.003-0.3$ & $0-1$\\
\hline 
 & $\Delta_1 \bar q$&$0.\pm0.02$&$0.01\pm0.04\pm0.03$&$0.01\pm0.04\pm0.03$\\
\hline
\end{tabular}

\vskip 1cm

Taking the first moments of valence distributions directly from the table,
one gets 
\be
\Delta_1 u_V -\Delta_1 d_V = 1.3\pm0.17\pm0.12,
\ee
and this result is in a good agreement with the equivalent of BSR (5)  
which within the unbroken sea approximation
is rewritten as
$$
\Delta_1 u_V-\Delta_1 d_V = \Delta_1 u-\Delta_1 d=\left|
 \frac{g_A}{g_V}\right| =1.2670\pm0.0035. 
$$

Let us now perform the similar analysis of  HERMES results for the first moments
of the polarized quark distributions published in Table 1 of ref. [8] which we,
for convenience, partially reproduce here

\vskip 0.3cm
\begin{tabular}{|c|c|c|c|}
\hline
 & Measured region& Low-x& Total integral\\
\hline 
$\Delta_1u+\Delta_1\bar u$& $0.51\pm0.02\pm0.03$ & $0.04$& $0.57\pm0.02\pm0.03$\\
$\Delta_1d+\Delta_1\bar d$& $-0.22\pm0.06\pm0.05$ & $-0.03$ & $-0.25\pm0.06\pm0.05$\\
$\Delta_1s+\Delta_1\bar s$& $-0.01\pm0.03\pm0.04$ & $0.00$ & $-0.01\pm0.03\pm0.04$\\
$\Delta_1\bar u$ & $-0.01\pm0.02\pm0.03$ & $0.00$ & $-0.01\pm0.02\pm0.03$\\
$\Delta_1\bar d$ & $-0.02\pm0.03\pm0.04$& $0.00$ & $-0.02\pm0.03\pm0.04$\\
$\Delta q_3$ & $0.74\pm0.07\pm0.06$& $0.07$ & $0.84\pm0.07\pm0.06$\\
$\Delta q_8$ & $0.32\pm0.09\pm0.10$ & $0.01$ & $0.32\pm0.09\pm0.10$\\
$\Delta_1 u_V$ & $0.52\pm0.05\pm0.08$ & $0.03$ & $0.57\pm0.05\pm0.08$\\
$\Delta_1 d_V$ & $-0.19\pm0.11\pm0.13$ & $-0.03$ & $-0.22\pm0.11\pm0.13$\\
\hline
\end{tabular}
\vskip 0.3cm

Directly from the table one gets
\be
\Delta q_3 \equiv (\Delta_1 u+\Delta_1\bar u)-(\Delta_1 d+\Delta_1\bar d)
= 0.82 \pm 0.06 \pm 0.06,
\ee
whereas the right-hand side ought to be equal to 
$|g_A/g_V| =1.2670\pm 0.0035 $ in accordance with the equivalent of BSR (3).

Thus, the HERMES distributions  do not satisfy the real equivalent of BSR (3)
(without any $Q^2$ dependence in the right-hand side). Instead these distributions  
are rather claimed to be 
in agreement with the  sum rule (see Eq. (13) of ref. [8])
$
\Delta q_3=\int_0^1\Delta q^{NS}dx=\left|g_A/g_V\right|\times C_{QCD}
$
(where 
$
\Delta q_{NS}(x,Q^2)\equiv \Delta u(x,Q^2)+\Delta\bar u(x,Q^2)- ((\Delta d(x,Q^2)+\Delta\bar d(x,Q^2)),
$
and
$C_{QCD}\equiv C_1^{NS}(Q^2)$ is the nonsinglet coefficient function\footnote{
The quantity 
$C_{QCD}$ in Eq. (13) of ref. [8] is namely the nonsinglet coefficient function
$C_1^{NS}$ given by Eq. (2) in $\rm 4^{th}$ order of QCD expansion, 
so that at $\alpha_s(2.5\,\,GeV^2)=0.35\pm0.04$ the right-hand side
of Eq. (13) in ref. [8]  reads
$C_1^{NS}|g_A/g_V|=1.01\pm0.05$ (just as in [8]).
For details see  [9], section 5.5.4, Eq. (5.22), Appendix A.7,
Eq. (A.44), and also [10], section 2.5 
} 
given by Eq. (2)) which is incorrect\footnote{Notice that the HERMES result (7) differs by about
$2$ standard deviations even from  this incorrect sum rule whose right-hand side reads
$ |g_A/g_V | \times C_1^{NS}(2.5\,\,GeV^2 ) =1.01\pm 0.05 $ (just as in [8]).}.

 To understand what happens let us  briefly remind the HERMES procedure
of the polarized density extraction from the measured SIDIS asymmetries. To this end
the method of purities is used at HERMES average $Q^2=2.5\, GeV^2$.
Within this method 
the leading order (LO) expression for SIDIS asymmetry 
\be
A_1^h(x,Q^2)=\frac{\sum_f e_f^2\Delta q_f(x,Q^2)\int_{0.2}^1dzD_f^h(z,Q^2)}  
{\sum_f e_f^2 q_f(x,Q^2)\int_{0.2}^1dzD_f^h(z,Q^2)}, \nonumber
\ee
is rewritten via purities $P_f^h(x,Q^2)$
as
\be
A_1^h(x,Q^2)=\sum_f\frac{\Delta q_f}{q_f}P_f^h,\quad
P_f^h(x,Q^2)\equiv\frac{e_f^2q_f(x,Q^2)\int_{0.2}^1dzD_f^h(z,Q^2)}
{
\sum_f e_f^2 q_f(x,Q^2)\int_{0.2}^1dzD_f^h(z,Q^2)
}, \nonumber
\ee
so that one can see that 
the application of the purity method
is equivalent to the leading order (LO) QCD analysis.  

Thus both SMC and HERMES collaborations use LO QCD analysis to extract
polarized distributions from the measured SIDIS asymmetries. However, 
there is important distinction between SMC and HERMES analysis conditions.
Namely,  whereas the SMC analysis
is performed at average $Q^2=10\, GeV^2$, i.e., when
LO QCD is a quite good approximation, 
the HERMES uses LO analysis    
to extract the polarized distributions from the respective asymmetries measured at
relatively  low average $Q^2 =2.5\, GeV^2$ value. So, the inconsistence of HERMES result
on $\Delta q_3$ with  the BSR 
can serve as a direct indication that LO analysis is not 
sufficient and NLO analysis is necessary at such conditions.   

It is illustrative to show how one can arrive at the incorrect sum rule
using the purity method at low average $Q^2$ value.

Since
the application of this method with respect to SIDIS asymmetries
is just LO QCD analysis, 
the first moments of the DIS structure functions  
$\Gamma_1^{p,n}$ have LO QCD expressions via HERMES distributions:
\be
\Gamma_1^{p}(2.5\, GeV^2) = \frac{1}{2}\sum_{q,\bar{q}} e^2_q\, 
\Delta_1 q(2.5\,\, GeV^2),\quad
\Gamma_1^{n}=\Gamma_1^{p}{\Bigl |}_{u\leftrightarrow d }.
\ee
On the other hand, the exact expression for the physical (independently 
measurable) 
quantity 
$\Gamma_1^{p}-\Gamma_1^{n}$ has a form (1), where 
$C_1^{NS}$ differs essentially from
the LO value $1$ at so low $Q^2$. \\
Now, if one equates (which is actually incorrect) the LO expression for 
$\Gamma_1^{p}-\Gamma_1^{n}$ derived from (8) to the exact expression (1), 
then one immediately obtains the sum rule (13) of ref. [8]. However, the 
quantities satisfying this sum rule certainly have nothing in common with 
the real LO first distribution moments (as well as with the real NLO, NNLO,... ones) 
which (as well as the real NLO, NNLO,... first moments) 
satisfy the equivalent of BSR (3) without any $Q^2$ dependence in the 
right-hand side.

In spite of it being almost obvious, it is expedient to show explicitly 
that the same trick, but in the NLO order (i.e., equating the quantity 
$\Gamma_1^{p}-\Gamma_1^{n}$, expressed via the NLO extracted distributions,  
to the exact value), 
would give rise to an error only of the order $O(\alpha_s^2)$. 

Indeed, the extraction of the quark distributions from the SIDIS asymmetries in NLO order means
that the respective DIS structure functions are expressed via these distributions as 
\be
g^p_1(x,Q^2)=\frac{1}{2}\sum_{q,\bar{q}} e^2_q\left( \Delta q
+\frac{\alpha_s(Q^2)}{2\pi}[\delta C_q\otimes
\Delta q+\delta C_g\otimes\Delta g]\right)(x,Q^2). \nonumber
\ee
 Then,  using  the explicit values of the first moments of the respective $\rm \overline{MS}$
Wilson coefficients [6] $M^1(\delta C_q)=-2,\quad M^1(\delta C_g)=0,$ one gets in NLO QCD:
\be
M^1[g_1^p] \equiv \Gamma_1^p= 
 \frac{1}{2}\sum_{q,\bar{q}} e^2_q\left(1-\frac{\alpha_s(Q^2)}{\pi}
 \right) \Delta_1 q,
\quad
\Gamma_1^{n}=\Gamma_1^{p}{\Bigl |}_{u\leftrightarrow d }.
\ee
Substituting this in the left-hand side of Eq. (1)
with $C_1^{NS}$ given by Eq. (2) reduced to NLO QCD: $C_1^{NS}=1-\alpha_s/\pi$,
one can see that $\alpha_s$ dependent multipliers
$\left (1-\alpha_s(Q^2)/\pi\right )$ cancel out precisely in the left- and right-hand 
sides, so that one arrives at 
Eq. (3) without any logarithmic corrections in the right-hand side.

On the other hand, 
setting the difference
$\Gamma_1^p -\Gamma_1^n$ composed from (9) equal to the left-hand side of (1) and  keeping (at will), 
simultaneously, 
in the right-hand side
the higher 
in $\alpha_s$ corrections (see Eq. (2)) for $C_1^{NS}$, one gets 
instead of  of BSR in a form (3) 
the sum rule with
$O\left(\alpha_s^2(Q^2)\right)$ terms in the right-hand side.

Let us now analyse the results of Table 1 of ref. [8] on $\Delta_1 \bar q$. 
First of all notice that HERMES uses the assumption that the relative polarization
of sea quarks is independent of flavour
\be
\frac{\Delta\bar u}{\bar u}=\frac{\Delta\bar d}{\bar d}=\frac{\Delta\bar s}{\bar s}=
\frac{\Delta s}{s},\nonumber
\ee
and this assumption is used to extract almost all first moments of the 
Table 1 of Ref. [8]\footnote{
Except for the quantity $\Delta q^*_8$ (see comment for the Table 1 of Ref. [8])
where the symmetric sea assumption
$
\Delta\bar u=\Delta\bar d=\Delta s=\Delta\bar s
$
is used.}.
It is of importance that this  assumption already 
implies the asymmetry of the light polarized sea quark distributions.
Indeed, the equality 
$\Delta \bar u/\bar u=\Delta\bar d/\bar d$, together with
the well known result\footnote
{
Notice that the equation $\bar u\neq \bar d$ is implicitly
used in [8]  since it is included in the parametrization CTEQ Low-$Q^2$
applied for the data analysis.
} [1] $\bar u(x)\neq \bar d(x)$ immediately give rise 
to $\Delta\bar u\neq \Delta\bar d$.  So, the results of Table 1 of Ref. [8]
for light sea quarks should be asymmetric. However, taking the first moments of the polarized light sea 
quark distributions directly from the Table 1, one gets
\be
\Delta_1\bar u-\Delta_1\bar d=(-0.01+0.02)\pm 0.061=0.01\pm 0.061,
\ee
which is just zero within the errors.

This disagreement now seems to be not too surprising
because the results of Table 1 of Ref. [8] do not satisfy the equivalents
of BSR  (3) and (5) (see discussion on Eq. (7)).

Let us now do some speculation assuming, for a moment, that at least
the first moments of the valence quark distributions from the Table 1 of
ref. [8]
are close to the real ones (satisfying the real equivalents of BSR (3) and (5)).  
Then, substituting  values $\Delta_1 u_V$ and $ \Delta_1 d_V$
taken from the Table 1 into
the BSR written in the form (5), one arrives at rather amazing result: 
\be
\Delta_1\bar u-\Delta_1\bar d=0.235\pm0.097,
\ee
i.e.,  the quantity 
$\Delta_1\bar u-\Delta_1\bar d$ we are interested in, 
is not zero as compared with the total error ($2.42$ standard deviations),
and, the polarized sea of light quarks is asymmetric
with respect to $u$ and $d$ quark polarized distributions.

Certainly, this is just a speculation based on the above-mentioned 
assumption. 
We rather believe that all this is a direct indication that the HERMES data
for asymmetries should be properly reanalysed. First, the low $x$ region
should be treated more carefully\footnote
{
Indeed, the unmeasured low-x region of HERMES is $0<x_{B}<0.023$, 
and in the all this rather large region HERMES uses the simple
Regge fit without the estimation of systematical errors.
}
and, second, the NLO QCD procedure is necessary
at so low $Q^2$ to properly extract  
so tiny quantities  as $\Delta_1 s$ and $\Delta_1 \bar u -\Delta_1\bar d$.

Besides, there is a good lesson here for another polarized SIDIS experiments,
in particular, for the
COMPASS experiment [11].
On the one hand the low $x_B$ boundary should be as small as possible
to achieve the maximal accuracy for the first moments.
On the other hand, 
it is extremely desirable to maximally increase the average $Q^2$ value in order to
the  simple LO analysis would become applicable. Otherwise, while the SIDIS
asymmetries are measured at average
$Q^2$ which is still about $2\,\, GeV^2$, the LO analysis 
is not sufficient and  NLO analysis is necessary to 
get reliable polarized distributions consistent
with the fundamental restrictions such as the Bjorken sum rule. 
     
\vskip 0.3cm

The authors are grateful to M.~Anselmino, R.~Bertini,
A.~Kataev, A.~Kotzinian,
A.~Maggiora, I.~Savin and O.~Teryaev  for
fruitful discussions.

\end{document}